\documentclass{article}
\usepackage[utf8]{inputenc}
\usepackage[english]{babel}
\usepackage{tikz}
\usepackage{pgfplots}
\usepackage{amsmath}
\usepackage{graphicx}
\usepackage{caption}
\usepackage{tikz}
\usepackage{tikz-3dplot} 
\usetikzlibrary{arrows,decorations.markings,math}
\usetikzlibrary{positioning,fit,backgrounds}

\usetikzlibrary{patterns.meta} 
\usepackage{caption}
\usepackage{subcaption}
\title{Freezing \textit{vs.} equilibration dynamics in the Potts model}
\author{Francesco Chippari and Marco Picco
\\}
\date{\textit{Sorbonne Universit\'e, CNRS UMR 7589, Laboratoire de Physique Th\'eorique et Hautes Energies, 4 Place Jussieu, 75252 Paris Cedex 05, France}
\\
\vspace{0.7cm}
\today}

\begin{document}

\maketitle
\begin{center}
    \section*{Abstract}
\end{center}
We study the quench dynamics of the $q$ Potts model on different bi/tri-dimensional lattice topologies.
In particular, we are interested in instantaneous quench from $T_i \rightarrow \infty$ to $T \leq T_s$, where $T_s$ is the (pseudo)-spinodal temperature.
The goal is to explain why,  in the large-$q$ limit, the low-temperature dynamics freezes on some lattices while on others 
the equilibrium configuration is easily reached. The cubic ($3d$) and the triangular ($2d$) lattices are analyzed in detail.
We show that the dynamics blocks when lattices have acyclic \textit{unitary structures} while the system goes to the equilibrium when 
these are cyclic, no matter the coordination number ($z$) of the particularly considered lattice.

\newpage
\section{Introduction, applications and motivations} 

The Potts model has been extensively studied by scientists for its wealth of features \cite{potts_1952,wu1982potts,baxter2016exactly}.
It is a generalization of the Ising model to the case in which the spin can take $q$ integer values.
The Ising model, being the milestone of statistical physics, shares its fame with him.
These two models have many common features and also some differences. 
In particular, both have a paramagnetic-ferromagnetic phase transition. But, while the transition is of $2^{nd}$ order in temperature for the Ising model, 
in dimensions two and three (the dimensions we will consider in this work), the Potts model becomes first order for $q > 4$ in two dimensions 
and $q > 2 $ in three dimensions.  The Potts model with large values of $q$ has been widely employed to analyse various phenomena
like the study of soap foams and cellular morphogenesis \cite{weaire1984soap, glazier1990coarsening, maree2007cellular}. 
Besides statistical physics, this model is widely used in many research domains such as, 
biophysics and bioinformatics to study protein evolution and folding \cite{10.1093/molbev/msab321,garel1988mean}, 
in quantitative social sciences and network theory \cite{bisconti2015reconstruction}, or high energy physics \cite{bass1999signatures} to 
study the qgp\footnote{Quark-gluon-plasma.}-hadron $1^{st}$ order phase transition in the scattering of heavy ions.

Over the last years, the dynamics of Potts model has also been the subject of many studies. Indeed, by varying the number of states $q$,  
one can explore different dynamical behaviours such as, metastability, coarsening, multinucleation and freezing 
\cite{Mazzarisi_2020, Chippari_2021, corberi2021many}. In \cite{Chippari_2021}, studying the dynamics through the $1^{st}$ order 
phase transition of the large-$q$ bidimensional Potts model, for the square and honeycomb lattice, we have found a peculiar freezing universal behaviour. 
At $T=0$ or $q \rightarrow \infty$ the dynamics is frozen forever and those blocked states are never escaped. Next, for finite sub-spinodal temperature and 
finite but still large $q$, after a universal time of the Arrhenius form $ t_s=e^{J/T}$, the equilibrium state is reached via coarsening. It was also found 
that the spinodal temperature $T_s$ below which the dynamics happens to be frozen up to $t_s$ is given by $T_s = 2 T_c / z$ with 
$z$ the coordination number of the lattice. On the contrary, for the triangular lattice  we do not see blocking at any temperature or $q$. 
Then a few natural questions are: why does the dynamics blocks for the square and honeycomb lattice but not for the triangular lattice? 
What are the lattice properties that induce these two kinds of behaviour? 

Following these previous results, it was suggested that the coordination number dictates the presence or absence of freezing. One of the aim 
of the present work is to test this suggestion by considering the cubic case which has the same coordination number as the bidimensional triangular lattice. 
In the following, we will show that freezing exists also for the cubic lattice, invalidating the claim that the coordination number characterises the low temperature
dynamics. Our results for the cubic lattice show the existence of a pseudo-spinodal temperature with the same expression, $T_s=  2 T_c / z$, 
with freezing for a quench below this temperature up to a universal time scale $e^{J/T}$. We will argue that it is in fact the \textit{unitary structure} of the 
lattices, which is made up of a central spin connected with its $z$ neighbours, called external spins, 
 which is responsible for the different types of behaviours at low temperatures. A lattice for which one can connect any pair of external spins by a path in 
 the unitary structure without going through the central spin is called  cyclic and otherwise it is called acyclic. 
For lattices with cyclic \textit{unitary structure}  there is no freezing, while if this is acyclic, blocked states will freeze the dynamics. 

The paper is organized as follows: in section 2 we recall some useful features of the Potts model and explain how we can determine
 the critical temperatures $T_c$  and the pseudo spinodal  temperatures $T_s$. In section \ref{d2d}, we recall previous results obtained 
 for the dynamical behaviour in two dimensions. Section \ref{d3d} contains our new results for the dynamics of the cubic lattice 
and in section \ref{cd3d} we give a physical argument for the different dynamical behaviours 
as a function of  the structures of the lattices. In section 6, we summarise our conclusions.

\section{The model, the algorithm and the low temperature region}
We are using the Potts model defined by:
%\\
%\vspace{0.3cm}
\begin{equation}
\label{eqn:Potts}
\displaystyle H_J\left[\left\{s_i\right\}\right]=\displaystyle- J \sum_{\langle i,j \rangle} \delta_{s_i,s_j}
\end{equation}
with $s_i \in \{1,2, \dots, q \}$, $i \in \{1, \dots, N=L^d\}$ and the sum is restricted to the nearest neighbours. 
In both two and three dimensions (that we will consider in this work), 
it has a first order paramagnetic-ferromagnetic phase transition for $q > 4$, \cite{wu1982potts}.
To study the out of equilibrium dynamics of our Potts model, various quenches to subcritical temperatures have been numerically simulated.
Since this model is not endowed with an intrinsic dynamics we have coupled it to a thermal reservoir
and we have used the \textit{Monte Carlo Markov-chain} technique to simulate a \textit{continuous time heat bath} 
stochastic dynamics \cite{BORTZ197510,newman1999monte,hassold1993fast}.
In $d=3$ dimensions, the critical temperature $T_c$ has not been determined rigorously analytically,
contrary to the $d=2$ case, \cite{baxter2016exactly}. One can use numerical simulation 
to determine $T_c$ as a function of $q$, for $d=3$. But, a simple argument, to determine it approximately,
can be obtained by comparing the probability of a paramagnetic and a ferromagnetic state. 
This simple argument gives a good approximation of the critical temperature in the case of $q \gg 1$, as we will check later on. 

\begin{figure}[!ht]
	\begin{center}
		\begin{tikzpicture}
		\draw[->] node[anchor=north west] {0} (0,0) -- (8,0);
%		\foreach \x/\xtext in {2/$\displaystyle 2T_c/z$}{
%			\draw (\x cm,-2pt) -- (\x cm,2pt) node[below] {\xtext};
%		}
%	\foreach \x/\xtext in {4/$T_c$}{
%			\draw (\x cm,-2pt) -- (\x cm,2pt) node[below] {\xtext};
%		}
		\node[label=$\hbox{disordered : }q^N$] at (6.0,0.0){};
		\node[label=$\hbox{ordered : }e^{\beta d N z/2}$] at (2.0,0.0){};
		\node[label= $T$] at (7.6,-0.8){};
		\draw (4.0 cm,-2pt) -- (4.0 cm,2pt);
		\node[label= $T_c$] at (4.0,-0.8){};
		\draw (2.0 cm,-2pt) -- (2.0 cm,2pt);
		\node[label= $2 T_c/z$] at (2.0,-0.8){};
		\end{tikzpicture}
	\end{center}
	\caption{Different phases and their total Boltzmann weight as a function of the temperature for the $d$ dimensional $q$ states Potts model.}
	\label{fig:sketch_Tc}
\end{figure}
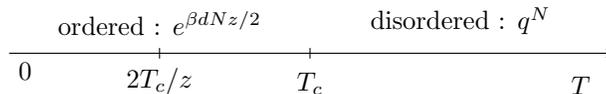

The argument goes as follow\footnote{$J=1$, from now on.}: at high temperature, the system is paramagnetic and completely disordered.
Thus, each disordered state has a probability of $\displaystyle P_{dis} \simeq \frac{q^N}{Z}$.
On the other hand, at low temperature, the system is ferromagnetic and ordered
with a probability, for each completely ordered state, of 
$\displaystyle P_{ord} \simeq \frac{e^{\beta N z/2}}{Z}$ with $z$ the coordination number of the lattice and $\beta=1/T$. 
A sketch is shown in Fig.~\ref{fig:sketch_Tc}. At $T_c$, these two probabilities should equalise, giving 
\begin{equation}
\label{eqn:eq_for_Tc}
\displaystyle e^{\beta_c N z/2}= q^N,
\end{equation}
and thus $\beta^*(q) \simeq \displaystyle (2/z)\log(q)$, where we use the index $\beta^*$ to differentiate it from the exact value $\beta_c$. 
Note that this argument is valid in any dimension. 

In $d=2$, $z=4$ for the square lattice, this simple argument 
predicts $\beta^*(q)  =\frac{1}{2} \log(q)$. The exact result, \cite{baxter2016exactly} is $\beta_c(q) = \log{(1+\sqrt{q})}\simeq \frac{1}{2} \log(q) + q^{-\frac{1}{2}} + \cdots$. 
Thus in the large $q$ limit, $\beta^*(q)$ goes to the exact result with a correction $O(1/\sqrt{q})$. We can also consider the case of the triangular lattice with $z=6$. 
Our prediction is $\beta^*(q) = \frac{1}{3} \log(q)$. The critical value is obtained by solving 
the equation: $x^3 -3 x +2-q  =0$, \cite{wu1982potts} with $x=e^{\beta_c}$. For $q > 4$, the solution is given by:
\begin{align}
	\displaystyle x = \displaystyle 2 \cosh\left\{\frac 23 \log\left(\frac{\sqrt{q}}{2}+\left(\frac{q}{4}-1\right)^{\frac{1}{2}}\right)\right\} \; .
\end{align}
In the large $q$ limit, the leading orders are 
$\beta_c(q) = \frac{1}{3} \log(q) + q^{-2/3} + O(q^{-1})$ which again converges toward $\beta^*(q)$. For the honeycomb lattice with $z=3$, 
one needs to solve the equation: $x^3 -3 x^2 -3 (q-1) x + 3 q -1 -q^2 =0$, \cite{wu1982potts}, for which 
we obtain $\beta_c(q) = \frac{2}{3} \log(q)  + \frac{\sqrt{5}-1}{2} q^{-1/3} + \cdots$, compared with $\beta^*(q) = \frac{2}{3} \log(q) $. Thus our 
naive argument gives a good approximation in $2d$ in the large $q$ limit. 

For the $3$ dimensional case, there are no predictions for the critical temperature to compare with. We have to rely on numerical measurements.
We will consider the cubic lattice for which $\beta^* = \frac{1}{3}\log(q)$ which becomes the critical inverse temperature in the large $q$ limit. 
\vspace{0.5cm}

\begin{figure} [!ht]
	\centering
		\centering
                 \includegraphics[width=\textwidth]{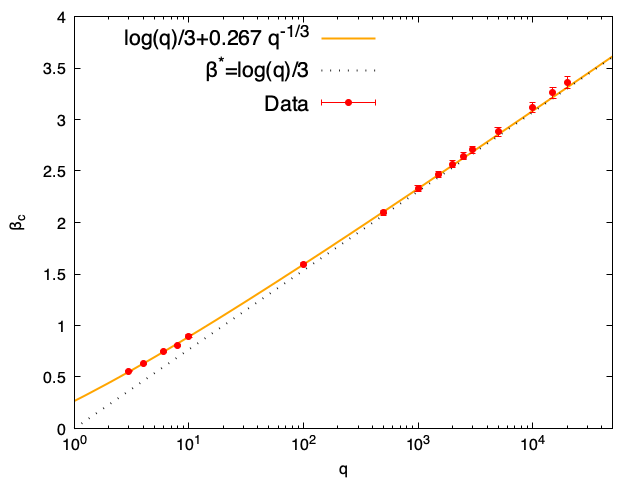}
	\caption{$\beta_c$ as a function of $q$ 
compared to $\beta^*$ and a best fit to the form 
	$\log{(q)}/3 + a q^{-1/3}$.}
	\label{fig:Tc_calculation}
\end{figure}

To determine numerically the position of the critical temperature for finite $q$ we proceed as follows. For a given value of $q$, we performed two simulations.
A first one in which we start from a low temperature and want to reach a disordered state, thus after having crossed the critical temperature. 
So we start with a completely ordered state at zero temperature. Next, we change the temperature to a high value $T_1$ and wait until a disordered state is 
obtained. We repeat this operation but with a smaller value for $T_1$. We call $T_c^+$ the smallest value for which we obtain a disordered state before some fixed 
large time that we chose as $t_{max}=20000 \times L^3$ updates. All our simulations were done using a non-local cluster algorithm, 
the \textit{Swendsen–Wang} \cite{wang1990cluster} which, close to the critical point, has a smaller autocorrelation time than local algorithms.
In the second simulation, we start from a high temperature and want to reach an ordered state. Starting from a completely disordered state, 
we quench it to a low temperature $T_2$ and wait until an ordered state is obtained. 
We repeat this operation while increasing $T_2$. The maximal value at which we observe an ordering (again in a time less than $t_{max}=20000 \times L^3$) is $T_c^-$. 
We then define the critical temperature as $T_c=(T_c^++T_c^-)/2$ 
with $\Delta T_c = \left |T_c^- - T_c^+\right| /2$.  Our simulations were performed on cubic lattices with periodic boundary conditions with size $L=32$ for $q \leq 10$ 
and $L=10$ for larger values of $q$. The results of our measurements are shown in Fig.~\ref{fig:Tc_calculation}.  We confirm the results obtained in the literature 
for $q=3$ and $q=4$ \cite{JANKE1997679,PhysRevB.40.5007}.  For large values of $q$ and for quench from high to low temperatures, metastability has to be considered \cite{Mazzarisi_2020} and this impacts the estimation of $T_c$. The obtained value for $T_c^-$ is pushed below the real $T_c$ and further as we increase the value of $q$. 
This explains why, in order to overcome the metastability effects, we have simulated only small sizes ($L=10$) for large values of $q$. 
In Fig.~\ref{fig:Tc_calculation}, we show, in fact, $\beta_c = 1 / T_c$ as a function of $q$ and we compare with the value $\beta^* = log{(q)}/3$. For large values of $q$, 
our measurements are close to this value, as expected. We still observe a small deviation which is due to the effect of metastability, 
as mentioned above. We also show a fit to the form $log{(q)}/3 + a q^{-b}$. We obtain an excellent fit to this form 
with $a=0.27 (1)$ and $b=0.35 (3)$. In Fig.~\ref{fig:Tc_calculation} and in the following of the paper,  we impose 
the value $b=1/3$  and then $a \simeq 0.267$. 

Thus we conclude that for large values of $q$, we have $T_c(q) \simeq 3/log(q)$. 
From the above fit, we obtain $T_c(q=100) \simeq 3/log(100)\times (1-0.0339)$, $T_c(q=1000) \simeq 3/log(1000)\times (1-0.0103)$ 
and $T_c(q=10000) \simeq 3/log(q)\times (1-0.0035)$. We will take into account these small deviations in the following. 

Last, we need to calculate the (pseudo)-spinodal temperature $T_s$. We use the word (pseudo)-spinodal since we are not in a mean-field
theory where an actual spinodal can be determined. This value establishes the boundary between the low temperature region and the metastable one. 
In the large $q$ limit, we start from a disordered state. During an update, one spin, belonging to a pair of neighbouring different spins, can flip such that a bond is created. This means that after the spin flip, the two neighbouring spins are in the same state. This increase the energy by $\beta$ and decreases the number of configurations by $q$. So, the weight $P_{bond}$ associated with a bond between 
these two spins is related to the weight of the two spins being different, $P_{dis}$, by: $P_{bond}=\displaystyle \frac{e^{\beta}}{q}P_{dis}$. 
Thus, if $e^{\beta} \gg q$, a bond will be created and it will be stable. On the contrary, for $q \ll e^{\beta}$, this bond will have a very small probability to be created and this leads to disordered metastable states.
This leads us to define $e^{\beta_s} = e^{J/T_s} = q$.  Then, using the relation $e^{\beta_c} = q^{2/z}$, from Eq.~(\ref{eqn:eq_for_Tc}), 
we deduce that $e^{\beta_s} =  e^{z \beta_c/2}$ and thereby,
\begin{equation}
\label{eqn:spinodal_3d_3}
\beta_s=\frac{z}{2}\beta_c \rightarrow T_s=\frac{2T_c}{z} \; .
\end{equation}
Thus, our low temperature region is delimited by an upper limit $T_s$ and is characterised by the fact that, when we quench the system in this region, 
it does not fall in metastable disordered states. Since for the square, honeycomb, triangular and even cubic lattices we know the critical 
temperature\footnote{For the square lattice, $T_c$ is known analytically for any $q$, \cite{baxter2016exactly}.} for large $q$, with a good approximation, 
we can easily construct the following table.

\begin{table}[!h]
	\centering
	{
		\begin{tabular}{ |c|c|c|c|c|} 
			\hline
			Lattice  & Dimensions & Coordination Number & $T_c(q \gg 1)$  & $T_s(q \gg 1)$ \\
			\hline
			Honeycomb &  $d=2$  & $z_h=3$& $3J/ 2\log(q)$ & $2T_c/3$\\
			\hline
			Square &  $d=2$  & $z_s=4$& $2J/ \log(q)$ & $2T_c/4$\\
			\hline		
			Triangular &  $d=2$  & $z_t=6$& $3J/ \log(q)$ & $2T_c/6$\\
			\hline
			Cubic &  $d=3$  & $z_c=6$ & $3J/ \log(q)$ & $2T_c/6$\\
			\hline
		\end{tabular}
	}
	\caption{Critical temperature and (pseudo)-spinodal for different lattice geometries in $d=2$ and $d=3$.}
	\label{Tab1}
\end{table}
Note that $T_s \sim 1/\log{(q)}$, in the large $q$ limit, for any lattice and dimension. 

\section{Dynamical behaviour in two dimensions}
\label{d2d}
We first recall some results for the dynamics of the Potts model in two dimensions. We will study in this work non-conserved dynamics \cite{bray2002theory} and consider 
the evolution of a system after a quench from a high temperature to a low temperature. The choice of the high temperature is not important, 
one usually starts from a completely disordered system at an infinite temperature. On the contrary, the choice of the final low temperature will 
influence the dynamical behaviour, as we will explain in the following. 

To describe the evolution of a system after a quench, we consider the growing length $R(t;q,T/T_c)$ 
defined as:
\begin{equation}
\label{Grow_leng}
	R(t;q,T/T_c)=\displaystyle {\frac{e_0(q,T/T_c)}{e_0(q,T/T_c)-e(t;q,T/T_c)}} \delta \; ,
\end{equation}
with $e(t;q,T/T_c)$ the density of energy at time $t$ for a given $q$ and $T$, $e_0(q,T/T_c)$ the 
equilibrium value \textit{i.e.}\ $e_0(q,T/T_c) = e(t\rightarrow \infty;q,T/T_c)$ and $\delta$ the lattice spacing.
For the sake of simplicity, we put $\delta=1$ in the following. After a quench at a low temperature, it happens that
the dynamics is able to bring the system to equilibrium via a coarsening process \cite{bray2002theory}. 
During this evolution, the system is described by a dynamical regime \cite{PhysRevB.33.7795,DERRIDA1997466} in which 
the growing length has a simple behaviour (for non-conserved dynamics):
\begin{equation}
	\label{coarsening}
    R(t) \sim t^{1/2} \text{ for $1 \ll R \ll L$} \; ,
\end{equation}
with $L$, the linear size of the considered system. This result is due to the dynamical scaling hypothesis \cite{bray2002theory,onuki2002phase,krapivsky_redner_ben-naim_2010} 
from which one deduces the independence of the dynamical exponent $z_d=1/2$ with respect to the parameters, $T/T_c$ and $q$. 

In practice, when our system is in the coarsening regime, the dynamics is ruled by the interfaces between clusters made of spins belonging 
to different states or, to make it simpler, having different colours. Indeed, the internal part of these domains is thermodynamically at equilibrium 
while the interfaces are creating a surface tension. These clusters tend to rearrange in order to minimise the free energy and this results in a process of 
interface diffusion or curvature-driven dynamics. In this regime, due to the dynamical scaling hypothesis, all the dynamical features can be described 
by using the growing length $R(t)$. 

The universal behaviour for $R(t) \sim t^{1/2}$ has been measured after a quench at a finite temperature for various values of $q$ and various types of lattices. Still, it was 
observed \cite{ferrero2007long,spirin2001fate, refId0} that at low temperature and for large values of $q$, freezing of the lattice occurs during very long times. In \cite{Chippari_2021}, 
a systematic study of this freezing was performed for the honeycomb lattice (with $z=3$), the square lattice ($z=4$) and the triangular lattice ($z=6$). 

For the honeycomb and the square lattice, it was shown that the dynamics is frozen, with a finite value of the growing length, after a quench at 
$T < T_s = \displaystyle \frac{2T_c}{z}$ and this, during a time $t_s  \simeq e^{1/T}$. Thus, for the square and honeycomb lattices, the dynamical behaviour can be expressed as
\begin{equation}
\label{eqn:sq_hon_dynamics}
\displaystyle R(t;T/T_c,q) \simeq f(e^{-1/T}t) \text{ with }  f(x) \simeq 
\begin{cases} \hbox{const}, & \mbox{for $x \ll 1$}  
\\ 
\displaystyle x^{1/2}, & \mbox{for $x \gg 1$}  \; .
\end{cases}
\end{equation}
Note that $R(t;T/T_c,q)$ remaining constant (and finite with a value $< L$) is the definition of a frozen state or blocked state (the two terms are used interchangeably).

\begin{figure}[!ht] %
	\centering
         \includegraphics[width=6cm]{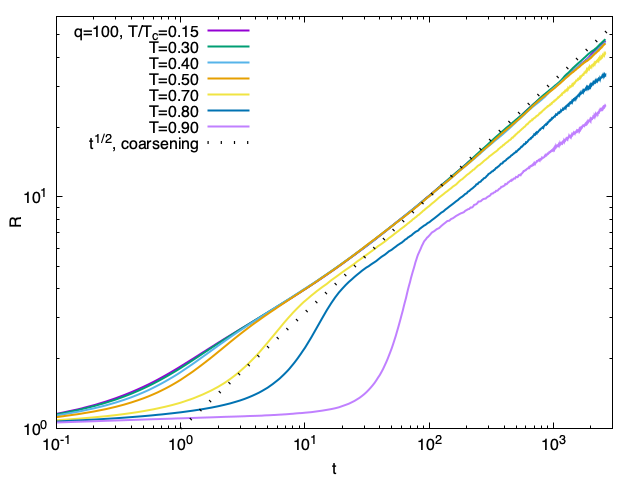}
         \includegraphics[width=6cm]{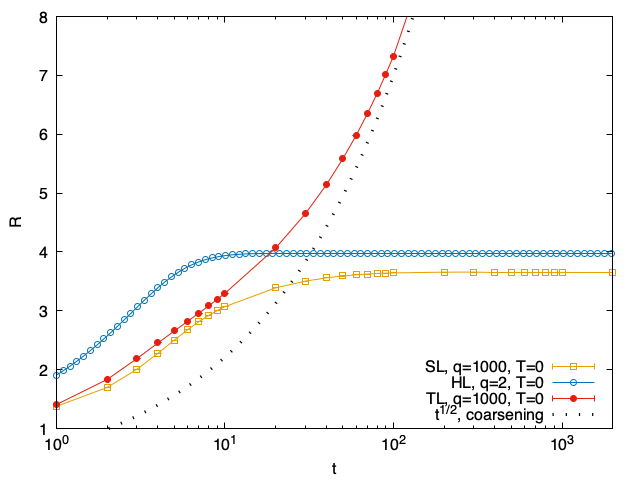}
 \caption{Left panel: Coarsening on the triangular lattice for $q=100$, $L=1000$ and different $T/T_c$ given in the key.
Right panel: Freezing on the square (SL) and honeycomb (HL) lattices and coarsening on the triangular one (TL), for $T=0$ and $q$ finite.
In both panels, the dashed line shows the standard coarsening $t^{1/2}$.}
 \label{fig:square2d}
     \hfill
\end{figure}

On the contrary, a simple coarsening is obtained for the triangular lattice with a coordination number $z=6$. 
This is shown in the left panel of Fig.~\ref{fig:square2d} in which we show $R(t)$ as a function of the time for various 
values of $T/T_c$, $q=100$ and $L=1000$  \cite{Chippari_2021}. We do not observe a change of behaviour as the quenched temperature 
crosses $T_s/T_c=1/3$.

%We have simulated the dynamics on the triangular lattice in a previous paper \cite{Chippari_2021}.
%Here we summarise these results with a simple plot, see Fig.~\ref{fig:tr2d} for $q=100$ and $L=1000$. 
% In that case, we expect $T_c= 0.63522$ and $T_s=0.21174$.
% 
%   \begin{figure}[!ht] %
%  	\centering
%   		\includegraphics[width=12cm]{Fig/tri.png}
%   	\caption{Coarsening on the triangular lattice for $q=100$, $L=1000$ and different $T/T_c$ given in the key. $T\leq T_s$ and $T > T_s$, \cite{Chippari_2021}.}
%   	\label{fig:tr2d}
%   	\hfill
%  \end{figure}
%The clusters are clearly not frozen and the curvature driven dynamics leads to equilibrium. 
%In the plot, we show also quench to $T > T_s$.  
%We do not observe a change of behaviour as we cross this value of $T_s$. 
The only difference is for quenches with $T$ close to $T_c$ for which longer and longer disordered metastable configurations survive having an $R(t)$ very close to $1$. A similar phenomenon exists also for the other types of lattices in two dimensions, 
the square and honeycomb lattices, \cite{Mazzarisi_2020}.  
But apart from this metastability, we do not observe a difference of behaviour in the coarsening for $T < T_c$.  Thus, there is no frozen configuration 
for the triangular lattice. Note that the fact that freezing is absent for the triangular lattice and present at $T < T_s$ on the square and 
honeycomb lattices was already observed in earlier works \cite{PhysRevB.28.2705,ANDERSON1984783,PhysRevA.43.2662}. In these works, 
it is also shown that the different behaviours are linked to topological considerations. 
%We will do a complementary analysis in the following.

The different types of behaviour for the growing length are shown for the three types of lattices in the right panel of Fig.~\ref{fig:square2d}, 
(SL denotes the square lattice,  HL the honeycomb lattice
and TL the triangular lattice). This figure shows results obtained in \cite{Chippari_2021} for the temporal behaviour of the growing length after a quench at zero temperature and 
for systems with linear size $L=1000$. For the triangular lattice, we just observe the standard coarsening. The same result is obtained for any value of $q$. 
For the honeycomb lattice, freezing is observed for any value $q \geq 2$ \cite{TakanoMiyashita,Chippari_2021}. For the square lattice, we also observe freezing. 
In \cite{Chippari_2021}, square lattices were considered only for $q \geq 100$. In fact, it can be checked that freezing is observed for any value of $q \geq 3$ for 
systems which are large enough. We computed the linear sizes $L_f(q)$ above which a quenched system remains always frozen at zero temperature.
In practice, we quenched $1000$ samples, each with a different initial random configuration, and waited up to $t=10^6$. $L_f(q)$ is the smallest linear size for which all the samples 
still have a growing length on the frozen plateau at $t=10^6$. We obtain $L_f(q=3) = 360, L_f(q=10) = 220$ and $L_f(q=100) = 30$. Note that this frozen state 
does not exist for $q=2$. One needs to have at least three colours in order to observe the freezing for the dynamics of the Potts model on the square lattice due to
the so-called  T-junctions \cite{glazier1990coarsening,lifshitz1962kinetics,Olejarz_2013}. 

Thus, freezing plays an important role in the dynamics with large $q$ or very small $T$ on the square and honeycomb lattices.
In this dynamical regime, featured by an $R(t)=\hbox{const}$, the dynamics is stuck for rather long time intervals, which become infinitely long 
when $q \rightarrow \infty$ or $T=0$. In \cite{Chippari_2021}, it was argued that the dynamics is the same in these two limits.
In this regime, the square lattice is decorated by some highly symmetric square and rectangular domains. Spins are in rather stable states 
and appropriate thermal fluctuations are needed to make them go into a state who triggers the coarsening process.
A detailed description of this regime is given in \cite{Chippari_2021}, the main 
information is that the thermal fluctuation needs to overcome an energy barrier of $e^{1/T}$, corresponding to the reversing of a corner, which explains 
the time scale $t_s  \simeq e^{1/T}$.

Last, in \cite{PhysRevB.28.2705, PhysRevB.33.7795,DERRIDA1997466, DERRIDA1996604}, the case with next to nearest
neighbours interactions in a square lattice topology has been considered and it was observed that the \textit{freezing} 
behaviour disappears for the Potts model for large $q$. Thus, increasing the coordination number from $z^{NN}=4$
to $z^{NNN}=8$ was belived to be another way to remove the freezing at low temperatures. But, as said previously, we will show that this is not the reason why freezing disappears.
 
We observe that the freezing exists for $z \leq 4$ and disappears for $z \geq 6$. In the following, we will consider the case 
of the cubic lattice in three dimensions which corresponds to $z=6$. We want to check the existence of freezing in this case.

\section{Dynamics on the cubic lattice} 
\label{d3d}
In this section, we present results for the evolution after quenches at low temperatures on the cubic lattice. In particular, we want to test if there exists, or not, 
a (pseudo)-spinodal temperature $T_s = 2 T_c/z$.  For earlier works on the cubic lattice, see \cite{doi:10.1080/13642818908220181}. 
We will consider quenches at various values of $T$ (small) such that:
\begin{align}
 T \leq T_s(q)=\frac{T_c(q)}{3}\simeq \frac{1}{log(q)+ 3 \times 0.267 q^{-1/3}} \; .
\end{align}
%
%% $\beta= log(q)/3 + 0.267 q^{-1/3}$
\begin{figure}[!ht]
\begin{center}
\includegraphics[width=12cm]{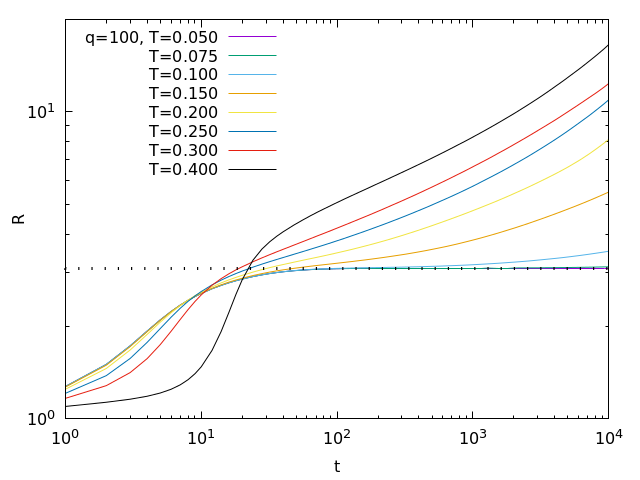}
\end{center}
\caption{$R(t)$ vs $t$ for $q=100$ and $L=400$. 
}
\label{FigRQ100}
\end{figure}

We first consider the case $q=100$ for which we have obtained $T_c(q=100) \simeq 0.6279$ and thus we expect $T_s(q=100) = 0.2093$. 
In Fig.~\ref{FigRQ100}, we show $R(t)$ as a function of the time $t$ after the quench at temperatures $0.05 \leq T \leq 0.4$ as shown in the caption. 
The data are for cubic lattice with a linear size $L=400$ and periodic boundary conditions. 
We checked that there are no finite size corrections to the behaviour of $R(t)$ except for $R(t) \simeq L$. 
At early times, up to $t\simeq 10$, we observe the same behaviour of $R(t)$  for all values of $T \leq 0.2 \simeq T_s(q=100)$. For this 
temperature range, $R(t)$ goes, at a later time, towards a plateau with value $\simeq 3.1$, as shown in the figure with a dashed line. $R(t)$ will 
remain on this plateau up to a time which increases while decreasing $T$. This indicates that, at small times, the dynamics for $T \leq T_s(q=100)$ is 
first described  by a blocked state corresponding to a zero temperature fixed point. The lower the temperature,
the longer it takes to escape from this blocked point. For $T > T_s(q=100)$, we observe that $R(t)$ is a function of the temperature for all the times 
and there is no blocked state.
\begin{figure}[!ht]
\begin{center}
\includegraphics[width=12cm]{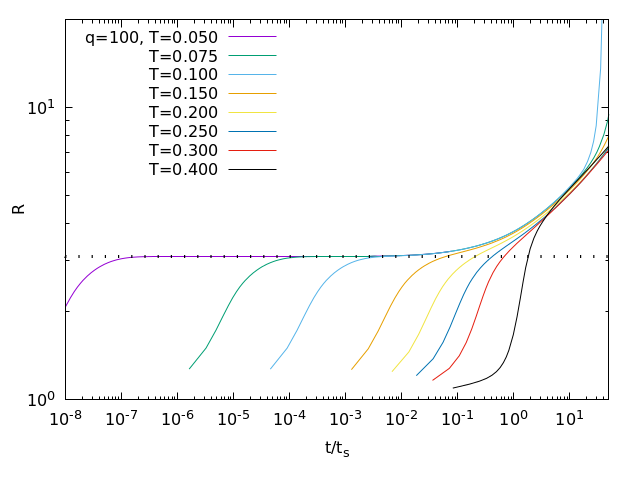}
\end{center}
\caption{$R(t)$ vs. $t/t_s$ for $q=100$ and $L=400$.
}
\label{FigRQ100b}
\end{figure}
In Fig.~\ref{FigRQ100b} we show $R(t)$ as a function of $t/t_s(T)$ with $t_s(T)=e^{1/T}$ a rescaling function (and the proper time of the dynamics in this regime). 
We observe that $R(t)$ escapes the plateau on a universal curve as a function of $t/t_s(T)$ for $T \leq T_s(q=100)$. This is similar 
to what was observed for the square lattice  in \cite{Chippari_2021}. 

\begin{figure}[!ht] %\ContinuedFloat	
\begin{center}
\includegraphics[width=12cm]{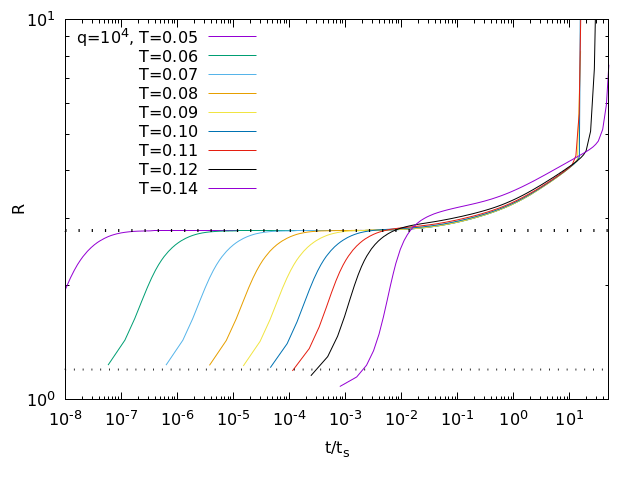}
\end{center}
\caption{$R(t)$ vs. $t/t_s$ for $L=400$ and $q=10^4$.
}
\label{FigRQ10k}
\end{figure}

Note that while the collapse after rescaling is obvious for small temperatures, it is less obvious, from Fig.~\ref{FigRQ100}, that the change of regime takes place 
exactly at $T = T_s$. This is because $q=10^2$ is not large enough. Next, we show in Fig.~\ref{FigRQ10k} similar data but for $q=10^4$ and for which  $T_c \simeq 0.3244$ and 
$T_s \simeq 0.1081$. In Fig.~\ref{FigRQ10k}, we observe a deviation starting for $T=0.110 > T_s$ and larger temperatures. 
Thus we expect that the change of behaviour at $T = T_s$ will become sharper as one increases $q$. This was also observed in $2d$, \cite{Chippari_2021}. 
This $q$ dependence also appears in the value of the growing length at the freezing plateau. We also repeated the same measurements for various values 
of $q$ ranging from $q=10$ up to $q=10^5$. We always observe the same behaviour for $q \geq 100$. (For $q=10$, the plateau is observed only at a very small temperature, up to $0.1$, while 
$T_s\simeq 0.4$). For the cubic lattice, we determined that the value of the plateau has a value $R(q) \simeq 2.774 + \hbox{const} \times q^{-3/4}$. This value corresponds to the freezing at zero temperature.

In summary, we observe blocking for the cubic lattice even if $z_{c}=z_t=6 > z_s$, see Fig.~\ref{FigRQ100}. The behaviour of $R(t)$ in this case, is 
very different from the one of the triangular lattice, see Fig.~(\ref{fig:square2d}). We have, thus, found that the energy barrier to be overcome 
to unleash the dynamics corresponds to a single spin-flip, which is the same as what is found in $d=2$ for the square lattice \cite{Chippari_2021}. 
Again, in the low temperature region it is necessary to wait a time which scales as the Arrhenius form $\sim e^{1/T}$ before being able to escape from a 
blocked configuration and equilibrate, see Fig.~\ref{FigRQ100b} and ~\ref{FigRQ10k}.

\section{Physical argument to freezing}
\label{cd3d}
We start from a configuration at an infinite temperature such that the lattice is completely disordered and spins are maximally uncorrelated. 
We consider the case of a general dimension and lattice. As already mentioned earlier, $T_s \simeq 1/\log{(q)}$ in the large $q$ limit. 
After an initial instantaneous quench at $T < T_s=1/\log{(q)}$, 
each update will change the value of a spin to be equal to one of its neighbours. 
Indeed, for any not-bonded spin, flipping to a value equal to a neighbouring spin will give a contribution to the free energy of $e^{1/T}$  larger than the lost contribution due to the entropy, $q$. 
After a time $t \simeq O(1)$, corresponding to the update of all the spins, a bond will be created among most spins and one of their neighbours chosen 
in a random way\footnote{At $t\simeq 1$ one expects $E \simeq - N/2 $ with $N$ the number of spins. Then $R(1) \simeq 1.2$. 
In Fig.~\ref{FigRQ10k}, we show as a thin dotted line this value. For each value of temperature smaller than $T_s \simeq 0.1081$, the first shown point for $R(t)$, 
corresponding to $t=1$, is above this line, while it is below for $T > T_s$. This is due to the metastability which occurs for these temperatures.}. 
At this stage, each spin is still completely unstable, even at $T=0$. It can change and create a bond with any of the others neighbours without an energy cost.
Then, these bonded spins, with the aim of minimising the free energy, 
create, eventually, other bonds. At this point, the dimension and the type of lattice become important, so we 
will consider different cases in the following. For simplicity, we will consider the case of a quench at $T=0$. 

\subsection{Frozen dynamics on the square lattice at $T=0$}
\begin{figure}[ht]
	\centering
	\begin{tikzpicture}[scale=1.3]
	\draw [color=red,mark=*] plot (-2,-1);
         \draw [color=red,mark=*] plot (-1,-1);	
         \draw [color=red,mark=*] plot (-1,-2);	
       	\draw [color=blue,mark=*] plot (0,-1);
         \draw [color=blue,mark=*] plot (1,-1);	
         \draw [color=blue,mark=*] plot (0,-2);	
	\draw [color=black!30!green,mark=*] plot (-2, 0);
	\draw [color=black!30!green,mark=*] plot (-1, 0);
	\draw [color=black!30!green,mark=*] plot (0, 0);
	\draw [color=black!30!green,mark=*] plot (1, 0);			
	\draw [color=black!30!green,mark=*] plot (-1, 1);
	\draw [color=black!30!green,mark=*] plot (0, 1);	
	\draw [ densely dotted] (0.1,0)--(0.9,0);
	\draw [ densely dotted] (-0.9,0)--(-0.1,0);
	\draw [ densely dotted] (-1.9,0)--(-1.1,0);		
	\draw [ densely dotted] (0.,0.1)--(0.,0.9);
	\draw [ densely dotted] (-1,0.1)--(-1,0.9);
	\draw [ densely dotted] (0.1,-1)--(0.9,-1);
	\draw [ densely dotted] (-1.9,-1)--(-1.1,-1);		
	\draw [ densely dotted] (0.,-1.9)--(0.,-1.1);
	\draw [ densely dotted] (-1,-1.9)--(-1,-1.1);
	\draw [  line width=0.5mm, color=black] (-2.5,-0.5)--(1.5,-0.5);	
	\draw [  line width=0.5mm, color=black] (-0.5,-0.5)--(-0.5,-2.5);		
		\end{tikzpicture}
		\caption{Frozen configuration at $T=0$ for the square lattice}
		\label{PFSq}
\end{figure}
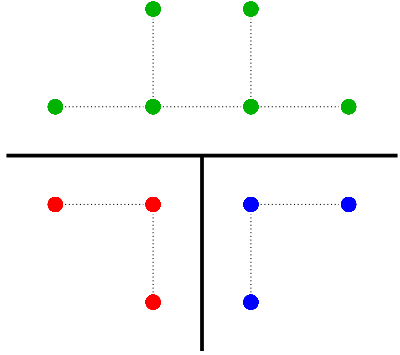

We first remind why the dynamics is frozen at $T=0$ on the square lattice for $q\geq 3$. 
This comes from the possibility of having particular configurations of spins as the one shown in Fig.~\ref{PFSq}, in which each colour corresponds to a different value of the spin. 
The four spins in the centre are completely stable. If we try to change the {\color{red} red} spin in a {\color{black!30!green} green} or {\color{blue} blue} spin, 
it would break two bonds while creating only one new bond. So, an increase of the energy which is forbidden at $T=0$. The same is true for 
the {\color{blue} blue} spin and for any of the two central {\color{black!30!green} green} spins,
changing colour would break three bonds and create only one new one. So, the four spins at the centre are frozen, it corresponds to the so-called $T$-junctions. 
Freezing for the square lattice at zero temperature is due to the existence of many such $T$-junctions and stable corners, see 
\cite{Olejarz_2013} for a more detailed description. 

 It is simple to check that similar corners exist on the honeycomb lattice leading also to freezing. For the honeycomb, the coordination number is $z=3$, in that case, two colours ($q=2$) are enough to freeze the lattice \cite{TakanoMiyashita}. 

\subsection{Not frozen dynamics on the triangular lattice at $T=0$}
\begin{figure}[!ht]
	\centering
	\begin{tikzpicture}[scale=1.5]	
	\draw [ densely dotted] (0,0)--(-1,0);
	\draw [ densely dotted] (0,0)--(-0.5,0.866);
	\draw [ line width=0.25mm, color=red] (0,0)--(-0.5,-0.866);
	\draw [ densely dotted] (0,0)--(1,0);
	\draw [ densely dotted] (0,0)--(0.5,0.866);
	\draw [ line width=0.25mm, color=red] (0,0)--(0.5,-0.866);
	\draw [ densely dotted] (0.5,-0.866)--(1,0);
	\draw [ densely dotted] (1,0)--(0.5,0.866);
	\draw [ densely dotted] (0.5,0.866)--(-0.5,0.866);
	\draw [ densely dotted] (-0.5,0.866)--(-1,0);
	\draw [ line width=0.25mm, color=red] (-0.5,-0.866)--(0.5,-0.866);
	\draw [ densely dotted]  (-1,0) -- (-0.5,-0.866);
	
	\draw [color=black!30!green,mark=*] plot (-1,0);
	\draw [color=blue,mark=*] plot (-0.5,0.866);
	\draw [color=red,mark=*] plot (-0.5,-0.866);
	\draw [color=magenta,mark=*] plot (1,0);
	\draw [color=orange,mark=*] plot (0.5,0.866);
	\draw [color=red,mark=*] plot (0.5,-0.866);
	\draw [color=red,mark=*] plot (0,0);
	\end{tikzpicture}
	\caption{Starting configuration for the triangular lattice}
	\label{FT1}
\end{figure}
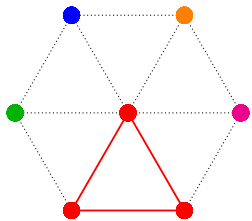

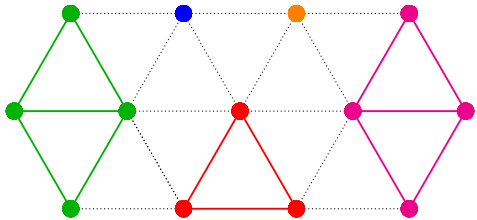
\begin{figure} [!ht]
	\centering
	\begin{tikzpicture}[scale=1.5]

	\draw [densely dotted ]  (-1,0) -- (-0.5,-0.866);
	\draw [ densely dotted] (0,0)--(-1,0);
	\draw [ densely dotted] (0,0)--(-0.5,0.866);
	\draw [ line width=0.25mm, color=red] (0,0)--(-0.5,-0.866);
	\draw [ densely dotted] (0,0)--(1,0);
	\draw [ densely dotted] (0,0)--(0.5,0.866);
	\draw [ line width=0.25mm, color=red] (0,0)--(0.5,-0.866);
	\draw [ densely dotted] (0.5,-0.866)--(1,0);
	\draw [ densely dotted] (1,0)--(0.5,0.866);
	\draw [ densely dotted] (0.5,0.866)--(-0.5,0.866);
	\draw [ densely dotted] (-0.5,0.866)--(-1,0);
	\draw [ line width=0.25mm, color=red] (-0.5,-0.866)--(0.5,-0.866);
	\draw [ densely dotted]  (-1,0) -- (-0.5,-0.866);
	\draw [ line width=0.25mm, color=black!30!green  ] (-2,0)--(-1,0);
	\draw [ line width=0.25mm, color=black!30!green ] (-2,0)--(-1.5,0.866);
	\draw [ line width=0.25mm, color=black!30!green  ] (-2,0)--(-1.5,-0.866);
	\draw [ line width=0.25mm, color=black!30!green  ] (-1.5,-0.866)--(-1,0);
	\draw [ line width=0.25mm, color=black!30!green  ] (-1,0)--(-1.5,0.866);
	\draw [ densely dotted  ] (-1.5,0.866)--(-0.5,0.866);
	\draw [ densely dotted  ] (-1.5,-0.866)--(-0.5,-0.866);
%	\draw [color=orange,mark=*] plot (1.5,0.866);
	\draw [ densely dotted  ] (1.5,0.866)--(1,0);
	\draw [ line width=0.25mm, color=magenta ] (2,0)--(1,0);
	\draw [ densely dotted  ] (2,0)--(1.5,0.866);
	\draw [ line width=0.25mm, color=magenta  ] (2,0)--(1.5,-0.866);
	\draw [ line width=0.25mm, color=magenta  ] (1,0)--(1.5,-0.866);	
	\draw [ line width=0.25mm, color=magenta  ] (2,0)--(1.5,0.866);
	\draw [ line width=0.25mm, color=magenta  ] (1,0)--(1.5,0.866);	
%	\draw [ line width=0.25mm, color=orange  ] (1.5,0.866)--(0.5,0.866);
    \draw [ densely dotted  ] (1.5,0.866)--(0.5,0.866);
	\draw [ densely dotted  ] (1.5,-0.866)--(0.5,-0.866);

	\draw [color=magenta,mark=*] plot (1.5,-0.866);
	\draw [color=magenta,mark=*] plot (1.5,0.866);	
	\draw [color=magenta,mark=*] plot (2,0);
	\draw [color=black!30!green,mark=*] plot (-1.5,0.866);
	\draw [color=black!30!green,mark=*] plot (-1.5,-0.866);
	\draw [color=black!30!green,mark=*] plot (-2,0);
	\draw [color=black!30!green,mark=*] plot (-1,0);
	\draw [color=blue,mark=*] plot (-0.5,0.866);
	\draw [color=red,mark=*] plot (-0.5,-0.866);
	\draw [color=magenta,mark=*] plot (1,0);
	\draw [color=orange,mark=*] plot (0.5,0.866);
	\draw [color=red,mark=*] plot (0.5,-0.866);
	\draw [color=red,mark=*] plot (0,0);
%	\draw [color=blue,mark=*] plot (-1,2*0.866);
%	\draw [color=orange,mark=*] plot (1,2*0.866);
%	\draw [color=blue,mark=*] plot (0,2*0.866);
%	\draw [line width=0.25mm, color=orange ] plot (+0.5,+0.866)--(+1,2*0.866);
%	\draw [densely dotted ] plot (+0.5,+0.866)--(0,2*0.866);
%	\draw [line width=0.25mm, color=orange] plot (1,2*0.866)--(+1.5,+0.866);
%	\draw [densely dotted ] plot (+1,2*0.866)--(0,2*0.866);
%	\draw [line width=0.25mm, color=blue ] plot (0,2*0.866)--(-0.5,+0.866);
%	\draw [line width=0.25mm, color=blue ] plot  (-1,2*0.866)--(-0.5,+0.866);
%	\draw [line width=0.25mm, color=blue ] plot (-1,2*0.866)--(0,2*0.866);
%	\draw [densely dotted ] plot (-1,2*0.866)--(-1.5,+0.866);
	\end{tikzpicture}
	\caption[Caption for the list of figures]{Intermediate configuration on the triangular lattice.}
%        \caption{$ $}
	\label{FT2}
\end{figure}

Next, we consider the case of  the triangular lattice at $T=0$ and check if similar blocked corners can exist or not. 
 To do so, let us start by considering hexagonal plaquettes, as shown in Fig.~\ref{FT1}.
Here the {\color{red}red} central spin has two bonds, \textit{i.e.}\ it has two neighbours with the same colour. We assume that the four 
other spins around this central one have different values (shown as different colours) and this in order to ensure that the {\color{red} red} spin can not flip to another value. 
We want to determine if it is possible to build a frozen configuration
with a fixed central {\color{red} red} spin. We first consider the {\color{black!30!green} green} spin on the left. This spin needs to be connected with three neighbours 
{\color{black!30!green} green} spins, as shown in Fig.~\ref{FT2}
to ensure that 
it can not  become {\color{red} red}. The same is true for the {\color{magenta} magenta} spin on the right. 
Then, one needs the configuration shown in Fig.~\ref{FT2}

\begin{figure} [!ht]
	\centering
	\begin{tikzpicture}[scale=1.5]

	\draw [densely dotted ]  (-1,0) -- (-0.5,-0.866);
	\draw [ densely dotted] (0,0)--(-1,0);
	\draw [ densely dotted] (0,0)--(-0.5,0.866);
	\draw [ line width=0.25mm, color=red] (0,0)--(-0.5,-0.866);
	\draw [ densely dotted] (0,0)--(1,0);
	\draw [ densely dotted] (0,0)--(0.5,0.866);
	\draw [ line width=0.25mm, color=red] (0,0)--(0.5,-0.866);
	\draw [ densely dotted] (0.5,-0.866)--(1,0);
	\draw [ densely dotted] (1,0)--(0.5,0.866);
	\draw [ densely dotted] (0.5,0.866)--(-0.5,0.866);
	\draw [ densely dotted] (-0.5,0.866)--(-1,0);
	\draw [ line width=0.25mm, color=red] (-0.5,-0.866)--(0.5,-0.866);
	\draw [ densely dotted]  (-1,0) -- (-0.5,-0.866);
	\draw [ line width=0.25mm, color=black!30!green  ] (-2,0)--(-1,0);
	\draw [ line width=0.25mm, color=black!30!green ] (-2,0)--(-1.5,0.866);
	\draw [ line width=0.25mm, color=black!30!green  ] (-2,0)--(-1.5,-0.866);
	\draw [ line width=0.25mm, color=black!30!green  ] (-1.5,-0.866)--(-1,0);
	\draw [ line width=0.25mm, color=black!30!green  ] (-1,0)--(-1.5,0.866);
	\draw [ densely dotted  ] (-1.5,0.866)--(-0.5,0.866);
	\draw [ densely dotted  ] (-1.5,-0.866)--(-0.5,-0.866);
%	\draw [color=orange,mark=*] plot (1.5,0.866);
	\draw [ densely dotted  ] (1.5,0.866)--(1,0);
	\draw [ line width=0.25mm, color=magenta ] (2,0)--(1,0);
	\draw [ densely dotted  ] (2,0)--(1.5,0.866);
	\draw [ line width=0.25mm, color=magenta  ] (2,0)--(1.5,-0.866);
	\draw [ line width=0.25mm, color=magenta  ] (1,0)--(1.5,-0.866);	
	\draw [ line width=0.25mm, color=magenta  ] (2,0)--(1.5,0.866);
	\draw [ line width=0.25mm, color=magenta  ] (1,0)--(1.5,0.866);	
%	\draw [ line width=0.25mm, color=orange  ] (1.5,0.866)--(0.5,0.866);
    \draw [ densely dotted  ] (1.5,0.866)--(0.5,0.866);
	\draw [ densely dotted  ] (1.5,-0.866)--(0.5,-0.866);
	\draw [line width=0.25mm, color=blue ] plot (0,2*0.866)--(-0.5,+0.866);
	\draw [line width=0.25mm, color=blue ] plot  (-1,2*0.866)--(-0.5,+0.866);
	\draw [line width=0.25mm, color=blue ] plot (-1,2*0.866)--(0,2*0.866);
	\draw [densely dotted ] plot (0,2*0.866)--(0.5,+0.866);
	\draw [densely dotted ] plot (-1,2*0.866)--(-1.5,+0.866);
	
	\draw [color=black!30!green,mark=*] plot (-1,0);
	\draw [color=blue,mark=*] plot (-0.5,0.866);
	\draw [color=red,mark=*] plot (-0.5,-0.866);
	\draw [color=magenta,mark=*] plot (1,0);
	\draw [color=orange,mark=*] plot (0.5,0.866);
	\draw [color=red,mark=*] plot (0.5,-0.866);
	\draw [color=red,mark=*] plot (0,0);
	\draw [color=black!30!green,mark=*] plot (-1.5,0.866);
	\draw [color=black!30!green,mark=*] plot (-1.5,-0.866);
	\draw [color=black!30!green,mark=*] plot (-2,0);
	\draw [color=magenta,mark=*] plot (1.5,-0.866);
	\draw [color=magenta,mark=*] plot (1.5,0.866);	
	\draw [color=magenta,mark=*] plot (2,0);
	\draw [color=blue,mark=*] plot (-1,2*0.866);
%	\draw [color=orange,mark=*] plot (1,2*0.866);
	\draw [color=blue,mark=*] plot (0,2*0.866);
%	\draw [line width=0.25mm, color=orange ] plot (+0.5,+0.866)--(+1,2*0.866);
%	\draw [densely dotted ] plot (+0.5,+0.866)--(0,2*0.866);
%	\draw [line width=0.25mm, color=orange] plot (1,2*0.866)--(+1.5,+0.866);
%	\draw [densely dotted ] plot (+1,2*0.866)--(0,2*0.866);	
	\end{tikzpicture}
	\caption[Caption for the list of figures]{Not blocked configuration on the triangular lattice.}
	\label{FT3}
\end{figure}
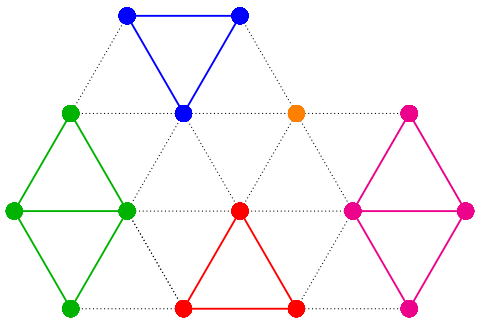

Next, we consider the {\color{blue} blue} spin. It needs to be connected with two other neighbouring spins which must be the two shown in Fig.~\ref{FT3}. 
Clearly, there is no way to make the {\color{orange} orange} spin stable since it can only have a bond with a single spin. 
Then, this {\color{orange} orange} spin will become {\color{blue} blue} or {\color{magenta} magenta}, allowing next the central {\color{red} red} spin to flip to the same colour.
This means that we can not build a blocked part of the configuration starting from the {\color{red} red} triangle. 

Note that it is still possible to obtain blocked configurations on the triangular lattice, but with much more complicated configurations 
involving many spins and occurring with a very small probability, as found in \cite{PhysRevE.99.062142}. One can argue that they require three 
different colours and produce the so-called three-hexagon state (with a small probability). But these structures have a very large growing length, a fraction of the linear size,
and are controlled by a time that scales as $L^2 \log{(L)}$. So, they do have no common point with the blocked states on the square lattice having a finite growing length reached
after a finite time.
\subsection{Frozen dynamics on the cubic lattice at $T=0$}
\begin{figure}[!ht]
\centering
\tdplotsetmaincoords{70}{120} % set viewpoint 
\tdplotsetrotatedcoords{0}{0}{0} %<- rotate around (z,y,z)
\begin{tikzpicture}[scale=2.55,tdplot_rotated_coords,
rotated axis/.style={->,purple,ultra thick},
spin/.style={ball color = black!30},
redspin/.style={ball color = red!1000},
greenspin/.style={ball color = green!1000},
bluespin/.style={ball color = blue!1000},
orangespin/.style={ball color = orange},
brownspin/.style={ball color = brown},
magentaspin/.style={ball color = magenta!1000}]
%ALL THE CUBE TO COLOR
%\foreach \x in {-1,0,1,2,3}
%\foreach \y in {0,1,2}
%\foreach \z in {0,1,2}{
%#####################################################
%\ifthenelse{  \lengthtest{\x pt < 2pt}  }{
%	\draw [densely dotted] (\x,\y,\z) -- (\x+2,\y,\z);
%	\shade[rotated axis,spin] (\x,\y,\z) circle (0.035cm); 
%}{}
%%#####################################################
%\ifthenelse{  \lengthtest{\y pt < 2pt}  }{
%	\draw [densely dotted]  (\x,\y,\z) -- (\x,\y+1,\z);
%	\shade[rotated axis,spin] (\x,\y,\z) circle (0.035cm);
%}{}
%%#####################################################
%\ifthenelse{  \lengthtest{\z pt < 2pt}  }{
%	\draw [densely dotted] (\x,\y,\z) -- (\x,\y,\z+1);
%	\shade[rotated axis,spin] (\x,\y,\z) circle (0.035cm);
%}{}
%
%}
\draw [line width=0.35mm, yellow!100 ] (1,1,1) -- (1,0,1);
\draw [line width=0.35mm, yellow!100 ] (1,1,1) -- (2,1,1);
\draw [line width=0.35mm, yellow!100] (1,1,1) -- (0,1,1);
\draw [line width=0.35mm, yellow!100 ] (1,1,1) -- (1,1,2);
\draw [line width=0.35mm, yellow!100 ] (1,1,1) -- (1,1,0);
JUST A CUBE
\foreach \x in {0,1,2}
\foreach \y in {0,1,2}
\foreach \z in {0,1,2}{
	%#####################################################
	\ifthenelse{  \lengthtest{\x pt < 2pt}  }{
		\draw [densely dotted] (\x,\y,\z) -- (\x+1,\y,\z);
		\shade[rotated axis,spin] (\x,\y,\z) circle (0.035cm); 
	}{}
	%#####################################################
	\ifthenelse{  \lengthtest{\y pt < 2pt}  }{
		\draw [densely dotted]  (\x,\y,\z) -- (\x,\y+1,\z);
		\shade[rotated axis,spin] (\x,\y,\z) circle (0.035cm);
	}{}
	%#####################################################
	\ifthenelse{  \lengthtest{\z pt < 2pt}  }{
		\draw [densely dotted] (\x,\y,\z) -- (\x,\y,\z+1);
		\shade[rotated axis,spin] (\x,\y,\z) circle (0.035cm);
	}{}
}
%red lines for equal spins
\draw [line width=0.25mm, red ] (1,1,1) -- (1,2,1);
\draw [line width=0.25mm, red ] (1,1,1) -- (2,1,1);
%central spin  
\shade[rotated axis,redspin] (1,1,1) circle (0.05cm);
%spins neig to c spherical form, like the grey ones.
\shade[rotated axis,greenspin] (1,0,1) circle (0.05cm);
\shade[rotated axis,redspin] (1,2,1) circle (0.05cm);
\shade[rotated axis,magentaspin] (0,1,1) circle (0.05cm);
\shade[rotated axis,redspin] (2,1,1) circle (0.05cm);
\shade[rotated axis,orangespin] (1,1,2) circle (0.05cm);
\shade[rotated axis,bluespin] (1,1,0) circle (0.05cm);
%\shade[rotated axis,greenspin] (2,0,1) circle (0.05cm);
%\shade[rotated axis,greenspin] (0,0,1) circle (0.05cm);
%\shade[rotated axis,orangespin] (1,2,2) circle (0.05cm);
%\shade[rotated axis,orangespin] (1,0,2) circle (0.05cm);
%\shade[rotated axis,magentaspin] (0,1,0) circle (0.05cm);
%\shade[rotated axis,magentaspin] (0,1,2) circle (0.05cm);
%\shade[rotated axis,bluespin] (1,0,0) circle (0.05cm);
%\shade[rotated axis,bluespin] (1,2,0) circle (0.05cm);
\end{tikzpicture}
	\caption[Caption for the list of figures]{\textit{Unitary structure} of the cubic lattice.}
	\label{FC1}
\end{figure}
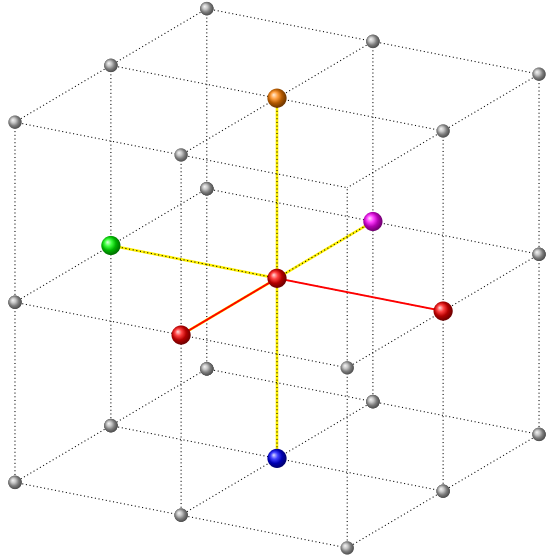

Let us now apply the very same argument but on the cubic lattice. As before, a one-bond kind of structure is useless for our aim, being fully unstable. 
Let us go directly to the interesting one. In Fig.~\ref{FC1}, we show how this lattice is made and its \textit{unitary structure}, made up of the highlighted 
yellow bonds plus the {\color{red} red} satisfied ones. For any lattice, by definition, the \textit{unitary structure} is made up of a central spin connected with 
its $z$ neighbours, called external spins. There exist two types of \textit{unitary structures}. The first type, which we call cyclic, for which one can connect any 
pair of external spins by a path in the \textit{unitary structure} without going through the central spin. The triangular lattice is cyclic. For the second type, 
which we call acyclic, the path needs to go through the central spin. The square lattice (with nearest neighbours), the hexagonal lattice and the cubic lattice are acyclic.

%%%%%%%%%%

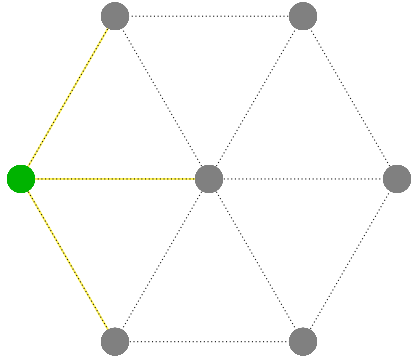
\begin{figure*}[!ht]
	\centering
		\begin{tikzpicture}[scale=2.5]			
			\draw [ line width=0.25mm, yellow!80] (0,0)--(-1,0);
			\draw [ line width=0.25mm, yellow!80]  (-1,0) -- (-0.5,-0.866);
			\draw [ line width=0.25mm, yellow!80] (-0.5,0.866)--(-1,0);
			\draw [ densely dotted] (0,0)--(1,0);
			\draw [ densely dotted] (0,0)--(0.5,0.866);
			\draw [ densely dotted] (0,0)--(0.5,-0.866);
			\draw [ densely dotted] (0.5,-0.866)--(1,0);
			\draw [ densely dotted] (1,0)--(0.5,0.866);
			\draw [ densely dotted] (0.5,0.866)--(-0.5,0.866);
			\draw [ densely dotted] (0,0)--(-1,0);
			\draw [ densely dotted] (0,0)--(-0.5,0.866);
			\draw [ densely dotted] (0,0)--(-0.5,-0.866);
			\draw [ densely dotted]  (-1,0) -- (-0.5,-0.866);
			\draw [ densely dotted] (-0.5,0.866)--(-1,0);
			\draw [ densely dotted] (-0.5,-0.866)--(0.5,-0.866);
			
			\draw [color=black!30!green,mark=*] plot (-1,0);
			\draw [color=gray,mark=*] plot (-0.5,0.866);
			\draw [color=gray,mark=*] plot (-0.5,-0.866);
			\draw [color=gray,mark=*] plot (1,0);
			\draw [color=gray,mark=*] plot (0.5,0.866);
			\draw [color=gray,mark=*] plot (0.5,-0.866);
			\draw [color=gray,mark=*] plot (0,0);
\end{tikzpicture}
\caption[Caption for the list of figures]{\textit{Unitary structure} of the triangular lattice.}
	\label{NFT}
\end{figure*}

In this configuration, again, the central spin has two neighbours with the same colour and four different neighbours. 
We consider that the {\color{black!30!green} green}, {\color{orange} orange}, {\color{magenta} magenta} and {\color{blue} blue} spins have already two satisfied bonds (not shown in Fig.~\ref{FC1} to avoid making the plot unreadable). 
They are all connected with the central one and two of them (the {\color{red} red}) have a satisfied bond. For each of these spins, we know three out of six neighbours.
Next, we consider one of the neighbours, say the {\color{black!30!green} green} one, and, 
as done in the triangular case, we look at its \textit{unitary structure}.
We know that it has two {\color{black!30!green} green} neighbours and a {\color{red} red} one. We assume that the other three neighbours 
have colours different from {\color{red} red} and at the maximum one neighbour which is {\color{orange} orange},
{\color{magenta} magenta} and {\color{blue} blue}. We remind also that we are in the limit of large $q$ and at an earlier time, so we expect that many different colours 
are present. Thus, the {\color{black!30!green} green} spin will remain {\color{black!30!green} green}. 
The exact same reasoning can be done for the other neighbours. We soon notice the common and the different features with respect to the triangular lattice.
Even if they share the same coordination number, here, the graph made by the \textit{unitary structure} has less edge and it is acyclic, 
while the triangular and hexagonal plaquette are cyclic.
Let us take, for example, the {\color{orange} orange} spin and let us watch the \textit{unitary structure} with the {\color{red} red} one in the centre.
We notice that the {\color{orange} orange} spin can only influence this structure by means of one eventual bond.
While, in the triangular lattice each spin at the vertex of a hexagon, including the ones not in the centre,  can influence the \textit{unitary structure} of a 
neighbouring hexagon by using up to three bonds, as shown in Fig.~\ref{NFT}. 

This difference is very important because this leads to the cyclicity of the \textit{unitary structure} and thus, to the mutual influence and connectivity among spins in it.
In the cubic lattice, indeed, external spins are not mutually influenced, as happens in the triangular one.
In fact, as shown previously, the blinking of a far away spin can lead to the unblocking of the dynamical behaviour. 
In the cubic case, instead, we do not have events able to activate the dynamics.
What we see is that soon, as in the square lattice case, the system reaches a blocked configuration in which highly regular rectangular, 
cubic, and square domains are present.
Here, so, a generalization of the results of \cite{Chippari_2021} holds.
Indeed, going to finite temperatures and finite $q$ we got exactly the same proper time as the one found in the square lattice.

Last, as we mentioned earlier, it was found \cite{PhysRevB.28.2705,PhysRevB.33.7795, DERRIDA1997466, DERRIDA1996604} that 
for the square lattice with next-to-nearest neighbours interactions, the \textit{freezing} behaviour disappears. It is simple to argue that with the 
addition of next to nearest neighbours the \textit{unitary structure} becomes cyclic for this lattice, explaining the absence of freezing.

\section{Conclusions}
In a previous work \cite{Chippari_2021}, it was observed that freezing exists at low temperatures, below the spinodal temperature $T_s$, for the $d=2$ square 
and honeycomb lattices, and for large $q$. The origin of this freezing is due to the existence  of blocked states at $T=0$. 
At finite temperature (and below $T_s$) the freezing is observed up to $t_s \simeq e^{J/T}$. This freezing was not observed for the triangular lattice 
which has a larger coordination number. In this work, we performed a similar study for the $3d$ cubic lattice (with the same coordination number) and found that freezing
exists at low temperatures. 
Next, with a mixture of numerical data and physical argument, we explain what leads to different low-temperature dynamical behaviours
 in the cubic and triangular lattices. We attribute this sharp change in the dynamics to the different topology of the lattice's \textit{unitary structures}.
In particular, when this is an acyclic graph, the spin-flips bring the system to a rather stable and highly symmetric blocked configuration, which is 
abandoned, thanks to thermal fluctuations, after an exponential time of the Arrhenius kind, $t_s=e^{1/T}$.
When, instead, the structure is cyclic, spins of the \textit{unitary structure} have a large "radius" of influence making impossible 
the blocking behaviour. This dynamics is, thus, of the coarsening kind even at very low temperature. 

\section{Acknowledgements}
F. Chippari wishes to thank F. Mori for useful and illuminating discussions.
\newpage
\bibliographystyle{ieeetr}
\bibliography{biblio}

\begin{thebibliography}{10}

\bibitem{potts_1952}
R.~B. Potts, ``Some generalized order-disorder transformations,'' {\em
  Mathematical Proceedings of the Cambridge Philosophical Society}, vol.~48,
  no.~1, p.~106–109, 1952.

\bibitem{wu1982potts}
F.-Y. Wu, ``The potts model,'' {\em Reviews of modern physics}, vol.~54, no.~1,
  p.~235, 1982.

\bibitem{baxter2016exactly}
R.~J. Baxter, {\em Exactly solved models in statistical mechanics}.
\newblock Elsevier, 2016.

\bibitem{weaire1984soap}
D.~Weaire and N.~Rivier, ``Soap, cells and statistics—random patterns in two
  dimensions,'' {\em Contemporary Physics}, vol.~25, no.~1, pp.~59--99, 1984.

\bibitem{glazier1990coarsening}
J.~A. Glazier, M.~P. Anderson, and G.~S. Grest, ``Coarsening in the
  two-dimensional soap froth and the large-q potts model: a detailed
  comparison,'' {\em Philosophical Magazine B}, vol.~62, no.~6, pp.~615--645,
  1990.

\bibitem{maree2007cellular}
A.~F. Mar{\'e}e, V.~A. Grieneisen, and P.~Hogeweg, ``The cellular potts model
  and biophysical properties of cells, tissues and morphogenesis,'' in {\em
  Single-cell-based models in biology and medicine}, pp.~107--136, Springer,
  2007.

\bibitem{10.1093/molbev/msab321}
M.~Bisardi, J.~Rodriguez-Rivas, F.~Zamponi, and M.~Weigt, ``{Modeling
  Sequence-Space Exploration and Emergence of Epistatic Signals in Protein
  Evolution},'' {\em Molecular Biology and Evolution}, vol.~39, 11 2021.
\newblock msab321.

\bibitem{garel1988mean}
T.~Garel and H.~Orland, ``Mean-field model for protein folding,'' {\em EPL
  (Europhysics Letters)}, vol.~6, no.~4, p.~307, 1988.

\bibitem{bisconti2015reconstruction}
C.~Bisconti, A.~Corallo, L.~Fortunato, A.~A. Gentile, A.~Massafra, and
  P.~Pell{\`e}, ``Reconstruction of a real world social network using the potts
  model and loopy belief propagation,'' {\em Frontiers in psychology}, vol.~6,
  p.~1698, 2015.

\bibitem{bass1999signatures}
S.~A. Bass, M.~Gyulassy, H.~Stoecker, and W.~Greiner, ``Signatures of
  quark-gluon plasma formation in high energy heavy-ion collisions: a critical
  review,'' {\em Journal of Physics G: Nuclear and Particle Physics}, vol.~25,
  no.~3, p.~R1, 1999.

\bibitem{Mazzarisi_2020}
O.~Mazzarisi, F.~Corberi, L.~F. Cugliandolo, and M.~Picco, ``Metastability in
  the potts model: exact results in the large q limit,'' {\em Journal of
  Statistical Mechanics: Theory and Experiment}, vol.~2020, p.~063214, Jul
  2020.

\bibitem{Chippari_2021}
F.~Chippari, L.~F. Cugliandolo, and M.~Picco, ``Low-temperature universal
  dynamics of the bidimensional potts model in the large q limit,'' {\em
  Journal of Statistical Mechanics: Theory and Experiment}, vol.~2021,
  p.~093201, sep 2021.

\bibitem{corberi2021many}
F.~Corberi, L.~F. Cugliandolo, M.~Esposito, O.~Mazzarisi, and M.~Picco, ``How
  many phases nucleate in the bidimensional potts model?,'' {\em Journal of
  Statistical Mechanics: Theory and Experiment}, vol.~2022, no.~7, p.~073204,
  2022.

\bibitem{BORTZ197510}
A.~Bortz, M.~Kalos, and J.~Lebowitz, ``A new algorithm for monte carlo
  simulation of ising spin systems,'' {\em Journal of Computational Physics},
  vol.~17, no.~1, pp.~10--18, 1975.

\bibitem{newman1999monte}
M.~E. Newman and G.~T. Barkema, {\em Monte Carlo methods in statistical
  physics}.
\newblock Clarendon Press, 1999.

\bibitem{hassold1993fast}
G.~N. Hassold and E.~A. Holm, ``A fast serial algorithm for the finite
  temperature quenched potts model,'' {\em Computers in Physics}, vol.~7,
  no.~1, pp.~97--107, 1993.

\bibitem{wang1990cluster}
J.-S. Wang and R.~H. Swendsen, ``Cluster monte carlo algorithms,'' {\em Physica
  A: Statistical Mechanics and its Applications}, vol.~167, no.~3,
  pp.~565--579, 1990.

\bibitem{JANKE1997679}
W.~Janke and R.~Villanova, ``Three-dimensional 3-state potts model revisited
  with new techniques,'' {\em Nuclear Physics B}, vol.~489, no.~3,
  pp.~679--696, 1997.

\bibitem{PhysRevB.40.5007}
C.-K. Hu and K.-S. Mak, ``Monte carlo study of the potts model on the square
  and the simple cubic lattices,'' {\em Phys. Rev. B}, vol.~40, pp.~5007--5014,
  Sep 1989.

\bibitem{bray2002theory}
A.~J. Bray, ``Theory of phase-ordering kinetics,'' {\em Advances in Physics},
  vol.~51, no.~2, pp.~481--587, 2002.

\bibitem{PhysRevB.33.7795}
J.~Vi\~nals and J.~D. Gunton, ``Fixed points and domain growth for the potts
  model,'' {\em Phys. Rev. B}, vol.~33, pp.~7795--7798, Jun 1986.

\bibitem{DERRIDA1997466}
B.~Derrida, ``Non-trivial exponents in coarsening phenomena,'' {\em Physica D:
  Nonlinear Phenomena}, vol.~103, no.~1, pp.~466--477, 1997.
\newblock Lattice Dynamics.

\bibitem{onuki2002phase}
A.~Onuki, {\em Phase transition dynamics}.
\newblock Cambridge University Press, 2002.

\bibitem{krapivsky_redner_ben-naim_2010}
P.~L. Krapivsky, S.~Redner, and E.~Ben-Naim, {\em A Kinetic View of Statistical
  Physics}.
\newblock Cambridge University Press, 2010.

\bibitem{ferrero2007long}
E.~E. Ferrero and S.~A. Cannas, ``Long-term ordering kinetics of the
  two-dimensional q-state potts model,'' {\em Physical Review E}, vol.~76,
  no.~3, p.~031108, 2007.

\bibitem{spirin2001fate}
V.~Spirin, P.~Krapivsky, and S.~Redner, ``Fate of zero-temperature ising
  ferromagnets,'' {\em Physical Review E}, vol.~63, no.~3, p.~036118, 2001.

\bibitem{refId0}
{Ib\'a\~nez de Berganza, M.}, {Ferrero, E. E.}, {Cannas, S. A.}, {Loreto, V.},
  and {Petri, A.}, ``Phase separation of the potts model in the square
  lattice,'' {\em Eur. Phys. J. Special Topics}, vol.~143, pp.~273--275, 2007.

\bibitem{PhysRevB.28.2705}
P.~S. Sahni, D.~J. Srolovitz, G.~S. Grest, M.~P. Anderson, and S.~A. Safran,
  ``Kinetics of ordering in two dimensions. ii. quenched systems,'' {\em Phys.
  Rev. B}, vol.~28, pp.~2705--2716, Sep 1983.

\bibitem{ANDERSON1984783}
M.~Anderson, D.~Srolovitz, G.~Grest, and P.~Sahni, ``Computer simulation of
  grain growth—i. kinetics,'' {\em Acta Metallurgica}, vol.~32, no.~5,
  pp.~783--791, 1984.

\bibitem{PhysRevA.43.2662}
E.~A. Holm, J.~A. Glazier, D.~J. Srolovitz, and G.~S. Grest, ``Effects of
  lattice anisotropy and temperature on domain growth in the two-dimensional
  potts model,'' {\em Phys. Rev. A}, vol.~43, pp.~2662--2668, Mar 1991.

\bibitem{TakanoMiyashita}
H.~Takano and S.~Miyashita, ``Ordering process in the kinetic ising mxodel on
  the honeycomb lattice,'' {\em Phys. Rev. B}, vol.~48, p.~7221, 1993.

\bibitem{lifshitz1962kinetics}
I.~Lifshitz, ``Kinetics of ordering during second-order phase transitions,''
  {\em Sov. Phys. JETP}, vol.~15, p.~939, 1962.

\bibitem{Olejarz_2013}
J.~Olejarz, P.~L. Krapivsky, and S.~Redner, ``Zero-temperature coarsening in
  the 2d potts model,'' {\em Journal of Statistical Mechanics: Theory and
  Experiment}, vol.~2013, p.~P06018, Jun 2013.

\bibitem{DERRIDA1996604}
B.~Derrida, P.~{de Oliveira}, and D.~Stauffer, ``Stable spins in the zero
  temperature spinodal decomposition of 2d potts models,'' {\em Physica A:
  Statistical Mechanics and its Applications}, vol.~224, no.~3, pp.~604--612,
  1996.

\bibitem{doi:10.1080/13642818908220181}
M.~P. Anderson, G.~S. Grest, and D.~J. Srolovitz, ``Computer simulation of
  normal grain growth in three dimensions,'' {\em Philosophical Magazine B},
  vol.~59, no.~3, pp.~293--329, 1989.

\bibitem{PhysRevE.99.062142}
J.~Denholm and S.~Redner, ``Topology-controlled potts coarsening,'' {\em Phys.
  Rev. E}, vol.~99, p.~062142, Jun 2019.

\end{thebibliography}
%\printbibliography
\end{document}